\documentclass[aps,prl,twocolumn,showpacs,groupedaddress,amsmath,amssymb,floatfix]{revtex4}
\usepackage{graphicx}

\begin{document}


\title{Dynamics of Internal Stresses and Scaling of Strain Recovery in an Aging Colloidal Gel}


\author{Ajay Singh Negi}
\email[]{ajay.negi@yale.edu}
\affiliation{Department of Chemical Engineering, Yale University,
New Haven CT 06511}
\author{Chinedum O. Osuji}
\email[]{chinedum.osuji@yale.edu}
\affiliation{Department of Chemical Engineering, Yale University,
New Haven CT 06511}.
\author{}
\affiliation{}


\date{\today}

\begin{abstract}
We monitor the relaxation of internal stresses in a fractal
colloidal gel on cessation of flow and find a weak power law decay,
$\sigma_i \sim t^{-\alpha}$ over 5 decades of time where $\alpha
\approx 0.07$. The system exhibits physical aging of the elastic
modulus, $G' \sim t^{\beta}$, with $\beta \approx \alpha$.
Imposition of zero stress after waiting time $t_w$ results in strain
recovery as the system relaxes without constraint. Remarkably,
recoveries at different $t_w$ can be shifted to construct a master
curve where data are scaled vertically by $1/\sigma_i(t_w)$ and
plotted horizontally as $(t-t_w)/t_w^{\mu}$ where $\mu\approx 1.25$,
indicative of a super-aging response.
\end{abstract}


\maketitle


Many out of equilibrium systems display a slow evolution of their
dynamics and properties in time, towards an eventual stationary or
equilibrium state. This process is referred to as ``aging'', and it
is the hallmark of many disordered materials, ranging from
molecular, polymer and spin glasses, to colloidal gels and glasses
\cite{Kob_PRL1997,McKenna_Angell2000,Hilgenfeldt2001,Cipelletti2002,Cipelletti_Farad2003}.
Although the origin of the aging response varies widely in these
systems, they share the commonality of a slow recovery of some
canonical structural or dynamic quantity after a departure from
equilibrium, usually initiated by a rapid change in a corresponding
control parameter. For example, after a sudden temperature quench in
polymer glasses, evolution of free volume and heat capacity lend
themselves to both conceptual and experimental investigation and
have proved highly successful as metrics for the recovery behavior
of polymers~\cite{ferry}. In spin glasses, below the spin glass
transition temperature $T_c$, removal of the external magnetic field
leads to a slow and non-exponential decay of remanent magnetization
on long time scales~\cite{binder}. In colloidal systems, the picture
is somewhat murkier \cite{mckenna_JRheo2009}. Attempts to correlate
the evolution in dynamics to measurable changes in structure have so
far been unsuccessful~\cite{cianci}. The concept of internal
strains, and the resultant stresses, due to out of equilibrium
locations of constituent matter is a potential starting point, for
systems with soft potentials. It has been advanced with some success
as a model for the dynamics of metallic glasses~\cite{egami} wherein
the distribution of stresses in the system defines the energy
landscape that is traversed in the slow return to equilibrium during
aging.

In this Letter, we consider strain recovery as a possible structural
metric for the aging of an attractive colloidal system and examine
the relationship between strain recovery and internal stress present
in the system. We study a dispersion which exhibits rapid gelation
via the formation of a fractal network on cessation of shear flow
and follow the evolution of the shear modulus in time after
gelation. We separately conduct experiments to monitor the
relaxation of internal stresses after flow cessation, as well as the
strain recovery of the system after various waiting times during the
stress relaxation. We find that the strain recovery behavior as a
function of system age can be collapsed into a single curve using a
vertical shift factor that is inversely proportional to the internal
stress at the start of the strain recovery and a horizontal shift
that scales inversely with elapsed time.

Our system consists of dilute dispersions, 2-6 wt.\%, of carbon
black particles (Cabot Vulcan XC72R) in a 3:1 mixture of mineral oil
and tetradecane (Aldrich Chemical Co.). Samples are prepared by
dispersing the particles in the solvent under vigourous vortexer
mixing for 2 minutes, followed by sonication for at least 30
minutes. Samples are studied in a stress controlled rheometer (Anton-Paar MCR301)
using a 50 mm 1$^{\circ}$ cone-plate geometry. The instrument is
mounted on an air table (Newport) to eliminate ambient mechanical vibrations
which could perturb the sample. Two separate sets of measurements were performed. The first was of the time evolution of the system modulus. The second characterized internal stress relaxation and subsequent strain recovery. The protocol for the modulus evolution consists of 4 steps: (1) a high shear rate rejuvenation step at $\dot\gamma$=1000 s$^{-1}$ which completely erases the flow history of the material and ensures a reproducible starting point \cite{COO:RheolActa2009}, (2) pre-shear at the
shear rate of interest, $\dot\gamma$=100 s$^{-1}$, for 1200 s to achieve a steady-state viscosity, (3) cessation of flow (the shear rate was reduced from 100 s$^{-1}$ to 0 in 0.1 s) and (4) measurement of the modulus as a function of time at $\omega=1$ rad/s and $\gamma=0.1$\%, within the linear viscoelastic regime. We set $t=0$ at the end of step 3, where flow stoppage leads to rapid system gelation. The protocol for the second set of experiments consists of five steps. Steps 1-3 are identical to that of the first protocol. In step 4, the sample is maintained under a quiescent or zero strain rate condition (the system is stationary).  The finite, decaying stress required to maintain the stationary condition is the internal stress \cite{osuji}. We measured the internal stress after cessation of flow as a function of time for a waiting period $t_w$. In step 5, at $t=t_w$, the zero strain rate is replaced by a zero stress condition, allowing the system to undergo strain recovery, analogous to that in traditional creep recovery experiments.  The recovered strain is measured as a function of time via the displacement of the rheometer tool. Measurements consisting of steps 1-5 were conducted for a range of $t_w$ from 1 to $10^4$s. The fourth step of our protocol provides a waiting time during which the sample ages quiescently. In typical rheological and dynamic scattering investigations, system properties are not monitored during the waiting step. The protocol applied here, however, permits correlation of the system properties measured after the waiting time with the internal stress dynamics characterized therein.

\begin{figure}[ht]
\includegraphics[width=80mm, scale=1]{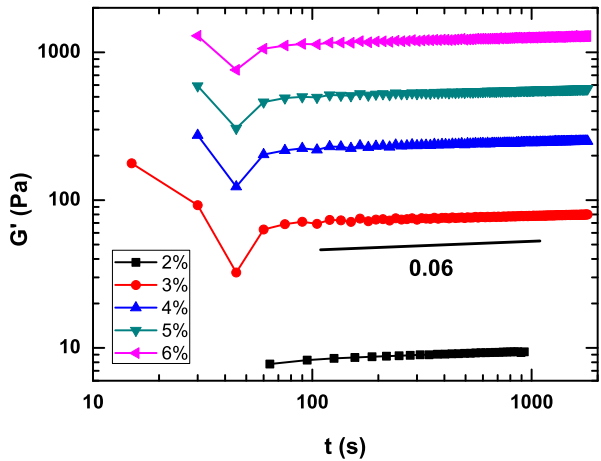}
\caption{Evolution of elastic modulus as measured via oscillatory shear with $\gamma$=0.1\% and $\omega$=1 rad/s.\label{modulus_aging}}
\end{figure}

Under the conditions of the experiment, the carbon black particles
interact via an attractive Van der Waals potential~\cite{trappe}
that results in the formation of a colloidal gel at rest. Gelation in
these systems is extremely fast~\cite{osuji,raghavan} such that a
substantial elastic modulus, well in excess of the viscous modulus,
is observed at the earliest measurable times after cessation of
flow, $\sim$ 0.1 s. The elastic modulus displays a slow power-law
increase with time, Figure~\ref{modulus_aging}, where
$G'\sim t^{\beta}$ with $\beta\approx 0.06$. This slow growth is due
to thermally driven structural evolution of the fractal gel, as
observed for analogous systems \cite{cipelletti}.

\begin{figure}[ht]
\includegraphics[width=80mm, scale=1]{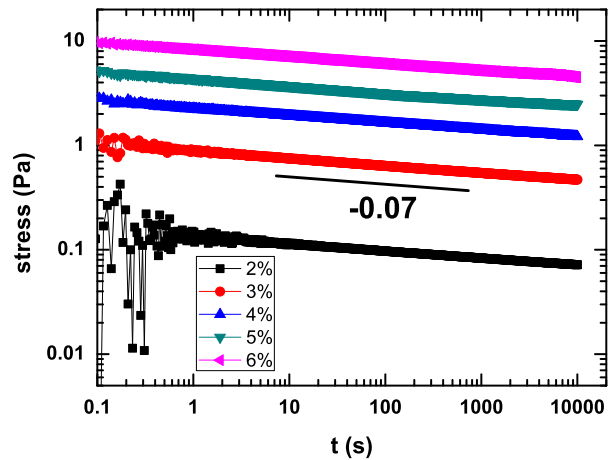}
\caption{Relaxation of internal stress as a function of time $t$ across
different samples studied.\label{stress_relax}}
\end{figure}

\begin{figure}[ht]
\includegraphics[width=80mm, scale=1]{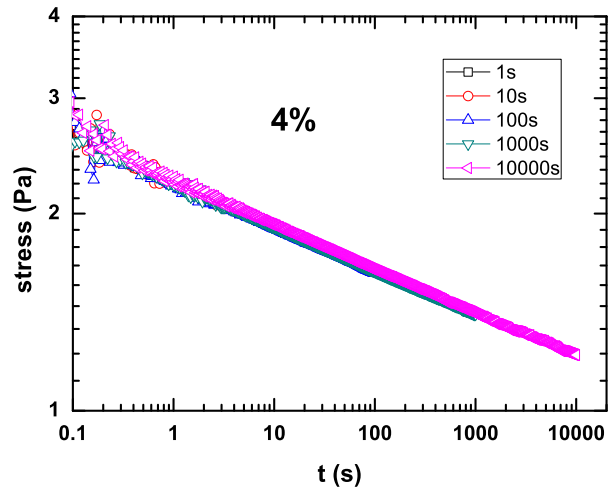}
\caption{Internal stress decay during different waiting times $t=t_w$ for
the 4 wt.\% sample. The stress is monitored up to the different values of $t_w$ shown in the legend in separate iterations of the experiment.\label{stress_relax_4}}
\end{figure}

Conversely, the internal stresses that are established in the system
during the rapid gelation on cessation of flow exhibit a weak decay
in time. The data are well fit by a power law, where the stress
decays as $\sigma_i\sim t^{-\alpha}$ over 5 decades of time with
$\alpha \approx 0.07$, as shown in Figure~\ref{stress_relax}. The
close correspondence between the scaling exponents of the stress
relaxation and the aging of the gel modulus suggests that the
relaxation of internal stresses may be a key indicator for the aging
of the system. Strain recoveries were recorded after various stress
relaxation durations, $t_w$. The trajectories of the stress decays
are well preserved across the numerous iterations of the stress
relaxation and strain recovery measurements, as shown for a 4 wt.\%
sample in Figure~\ref{stress_relax_4}. Similar consistent behavior
was observed for the other concentrations underlining the efficacy
of the rejuvenating step in effectively eradicating the flow history
of the sample.

\begin{figure}[ht]
\includegraphics[width=80mm, scale=1]{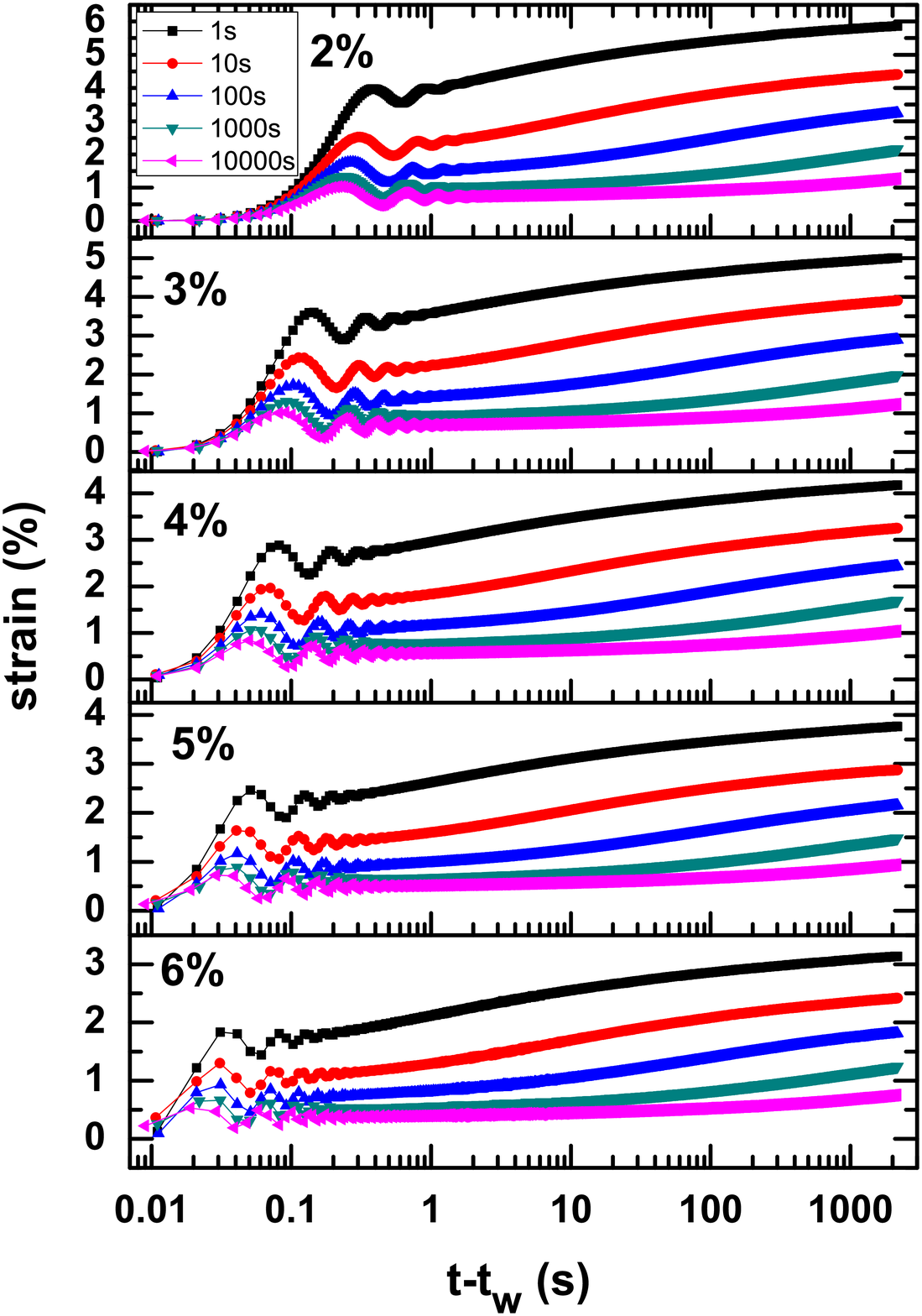}
\caption{Strain recovery after pre-shear at 100 s$^{-1}$ for different samples after different waiting times $t_w$. The concentration is mentioned in the figure. The times mentioned in the legend are different values of $t_w$.\label{strain_recovery_single}}
\end{figure}

\begin{figure}[ht]
\includegraphics[width=80mm, scale=1]{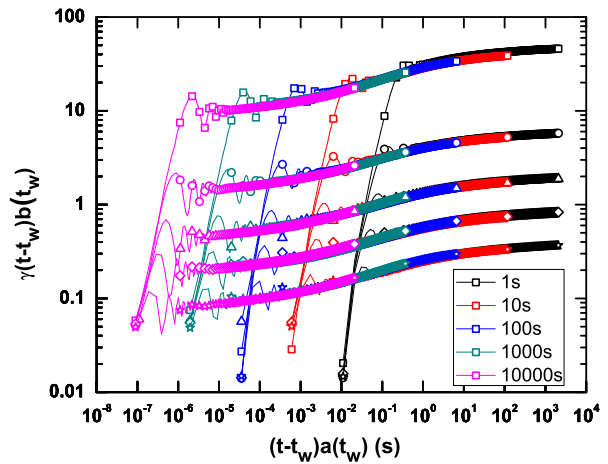}
\caption{Strain recovery master curves assembled from data taken as
a function of $t_w$. Squares, circles, triangles, diamonds and stars
denote 2 wt.\%, 3 wt.\%, 4 wt.\%, 5 wt.\% and 6 wt.\% concentrations
respectively. $a(t_w)=t_w^{-5/4}$, $b(t_w)=\sigma_i(t_w)^{-1}$. The times mentioned in the legend are different values of $t_w$.
\label{strain_recovery_master}}
\end{figure}

The imposition of zero stress after waiting times $t_w$ of  1, 10, 100,
1E3 and 1E4 s resulted in strain recoveries ranging from $1-5$\%,
measured over 2000s, Figure~\ref{strain_recovery_single}. At very
short times, oscillations are present in the data due to inertially
driven ringing of the sample. Such ``creep ringing'' is commonly
observed in elastic gels when subjected to a stress impulse~\cite{baravian}. Here,
the impulse originates from the internal stress of the sample itself
which drives an initial fast recoil of the tool that is quickly
damped out by the viscosity of the sample.
Thereafter, the material undergoes a slow strain recovery as
structural rearrangements occur during aging. Younger samples, those
at smaller $t_w$, display an initial rapid strain recovery which
appears to asymptote towards a long time plateau value. By contrast
older samples exhibit slower initial recovery from a short time
plateau, followed by a gradual rise at long time. Samples of
intermediate age show an inflection in their recovery. Remarkably,
strain recovery data from different sample ages can be shifted to
construct a single master curve at each composition,
Figure~\ref{strain_recovery_master}.

The vertical shift factors are simply inversely proportional to the
value of the internal stress at the start of the strain recovery,
$b(t_w)=1/\sigma_i(t_w)$. The horizontal shift factor
$a(t_w)=t_w^{-\mu}$ where the best overlay is obtained for
$\mu=5/4$. The vertical shift of the data by the inverse of the
internal stress is strikingly simple. It is representative of an
elastic strain that is recovered quickly at short times upon the
start of the measurement. At long times, there is a far slower and
continuous recovery of strain that persists out to the longest times
measured, over $1000$ s. We have ruled out external influences such
as tool drift or room vibration in the system. It should be noted
that the complex modulus and thus the dynamic viscosities of the
samples are large, $>200$ Pa for the data of Figure
\ref{strain_recovery_single}, so any small perturbations from
external sources would be quickly dissipated. The shifting of
dynamical responses for older samples to short rescaled time is
commonly encountered in soft glassy systems
\cite{Cloitre_PRL2000,shi,ovarlez,joshi} and has been related
phenomenologically to the elapsed time rescaling first successfully
advanced for polymer melts \cite{struik}. In these systems, however,
either simple aging ($\mu=1$) or sub-aging ($\mu < 1$) are observed,
whereas here we obtain the best overlap for $\mu>1$, indicative of a
super-aging response. The master curves indicate that the dynamics
of strain recovery at long times should be understood from the
behavior of young materials, and vice-versa, that the dynamics at
short times are exhibited within the observation window by older
samples. Such a display is counter intuitive if viewed simply in the
context of recovered strain as marking the proximity to long-time
equilibrium. In that sense, older samples would show more asymptotic
tendency than younger samples and not the other way around as
observed. This display, however, depends on the duration of the
waiting time relative to the timescale for complete relaxation. For
short $t_w$, we can view the response purely as a reflection of the
slow vs. fast dynamics of old vs. young samples. Thus we expect
elapsed time superposition to hold as it indeed does. The near
uniform scaling of the internal stress relaxation with time across
the different compositions studied (Figure \ref{stress_relax},
$\left<\alpha\right> = 0.07 \pm 0.002$) enables the strain recovery
to be simply scaled in terms of waiting time alone where the
vertical shift factor is now $b\sim t_w^{\alpha}$. The data of
Figure \ref{strain_recovery_master}, across 5 compositions, can all
be rescaled with the same shift factors with very good fidelity.

Internal stresses are known to persist for long times in a variety
of disordered systems~\cite{ramos,hartley,bellour,ranjini}.
Stochastic, local stress relaxation events have been invoked as a
concept to explain the unusual intermittency and ballistic dynamics
observed in various soft materials~\cite{ramos,bellour,ranjini}. The
origin of these stresses is in the non-equilibrium location of
system constituents. In colloidal systems with soft potentials, the
sudden nature of an ergodicity breaking transition results in the
arrest of particles away from preferred locations, that is those
where the gradient of the inter-particle potential is minimized. As
a result, forces exist among particles locally, and are propagated
throughout the system along particle contact chains or branches of
the fractal network in the case of gels. The departure of the system
from equilibrium is encoded by the distribution of local
displacements of particles with respect to their equilibrium
positions. Aging occurs via structural rearrangements that minimize
these internal stresses as particles conduct a thermally driven
exploration of their local potential energy landscape. In this
framework, it is clear that the evolution of local strain, and in
response, local stress, chart the aging process. In a gel which is
formed by the slow aggregation of freely diffusing species, such
local displacements and stresses are directionally random and the
resultant macroscopically observable stress is zero, or vanishingly
small. In the present system, the application of shear implies that
the displacement of particles from their preferred equilibria and
the deformation of particle clusters will be biased along the flow
direction. The rapid gelation that occurs on cessation of flow then
results in residual stresses that are macroscopically non-zero. As
expected, the stress acts counter to the direction of the
deformation, so the sign of the stress is reversed if the direction
of the shear flow is changed \cite{osuji}.

The correspondence between the relaxation of internal stress and the
increase of the gel modulus point strongly to the identification of
the internal stress state as an indicator for the aging response of
the system. From a microscopic perspective, due to the deforming
effect of the shear flow before gelation, structural rearrangements
involved in aging are anisotropic and thus give rise to
macroscopically observable strain recovery. The dynamics of this
strain recovery then serve as a well rationalized metric of the
aging behavior of the system, as demonstrated here. At long times,
the total recovered strain, $[\gamma(t-t_w)]_{t=\infty}$, should be
smaller for older samples, which have undergone longer stress
relaxation under zero strain. The master curves however imply no
significant difference in the long-time asymptotic values of
recovered strain for different sample ages. This is an implication
of the smallness of the experimental time scales relative to that
for the full aging of the system. This point is underscored by the
continuous power-law dependence of the internal stress relaxation
over 5 decades of time.

\begin{acknowledgments}
The authors gratefully acknowledge L. Cipelletti, V. Trappe and D. Weitz for  helpful discussions and NSF funding via CBET-0828905.
\end{acknowledgments}

\bibliography{scaling_strain_recovery_gel_Ajay}

\end{document}